\documentclass{ws-procs975x65}

\def\mnras{Mon. Not. Roy. Astr. Soc.}
\def\apj{Astrophys. J.}
\def\apjl{Astrophys. J. Lett.}

\def\aapr{Astron. Astrophys. Rev.}

\def\prc{Phys. Rev. C}

\begin{document}

\title{SGRS AND AXPS AS MASSIVE FAST ROTATING HIGHLY MAGNETIZED WHITE DWARFS: THE CASE OF SGR 0418+5729}

\author{K. BOSHKAYEV$^{1,2,3}$, J.A. RUEDA$^{1,2}$ AND R. RUFFINI$^{1,2}$}

\address{$^1$Department of Physics and ICRA, Sapienza University of Rome,\\
Aldo Moro Square 5, I-00185 Rome, Italy\\
$^2$ICRANet, Square of Republic 10, I-65122 Pescara, Italy\\
$^3$Physicotechnical Department, Al-Farabi Kazakh National University,\\
Al-Farabi avenue 71, 050038 Almaty, Kazakhstan;\\
$^*$E-mail: kuantay@icra.it, jorge.rueda@icra.it, ruffini@icra.it}

%\author{A. N. AUTHOR}
%
%\address{Group, Laboratory, Street,\\
%City, State ZIP/Zone, Country\\
%E-mail: an\_author@laboratory.com}

\begin{abstract}
We describe one of the so-called {\it low magnetic field magnetars} SGR 0418+5729, as a massive fast rotating highly magnetized white dwarf following Malheiro et. al.\cite{M2012}. We give bounds for the mass, radius, moment of inertia, and magnetic field for these sources, by requesting the stability of realistic general relativistic uniformly rotating configurations. Based on these parameters, we improve the theoretical prediction of the lower limit of the spin-down rate of SGR 0418+5729. In addition, we compute the electron cyclotron frequencies corresponding to the predicted surface magnetic fields.
\end{abstract}
\keywords{general relativistic white dwarfs; SGRs and AXPs; spin-down rate.}

\bodymatter
%%%%%%%%%%%%%%%%%%%%%%%%%%%%%%%%%%%%%%%%%%%%%%%%%%%%%%%%%%%%%%%%%%
%%%%%%%%%%%%%%%%%%%%%%%%%%%%%%%%%%%%%%%%%%%%%%%%%%%%%%%%%%%%%%%%%%
\section{Introduction}\label{sec:1}
%%%%%%%%%%%%%%%%%%%%%%%%%%%%%%%%%%%%%%%%%%%%%%%%%%%%%%%%%%%%%%%%%%
%%%%%%%%%%%%%%%%%%%%%%%%%%%%%%%%%%%%%%%%%%%%%%%%%%%%%%%%%%%%%%%%%%

Soft Gamma Ray Repeaters (SGRs) and Anomalous X-ray Pulsars (AXPs) are a class of compact objects that show interesting observational properties see e.g. Mereghetti (2008)\cite{mereghetti08}: rotational periods in the range $P\sim (2$--$12)$ s, a narrow range with respect to the wide range of ordinary pulsars $P\sim (0.001$--$10)$ s; spin-down rates $\dot{P} \sim (10^{-13}$--$10^{-10})$, larger than ordinary pulsars $\dot{P} \sim 10^{-15}$; strong outburst of energies $\sim (10^{41}$--$10^{43})$ erg, and for the case of SGRs, giant flares of even large energies $\sim (10^{44}$--$10^{47})$ erg, not observed in ordinary pulsars.

The observation of SGR 0418+5729 with a rotational period of $P=9.08$ s, an upper limit of the first time derivative of the rotational period $\dot{P} < 6.0 \times 10^{-15}$ Rea et. al.(2010)\cite{rea10}, and an X-ray luminosity of $L_X = 6.2\times 10^{31}$ erg s$^{-1}$ can be considered as the Rosetta Stone for alternative models of SGRs and AXPs.

The magnetar model, based on a neutron star of fiducial parameters $M=1.4 M_\odot$, $R=10$ km and a moment of inertia $I = 10^{45}$ g cm$^2$, needs a magnetic field larger than the critical field for vacuum polarization $B_c=m^2_e c^3/(e \hbar)=4.4\times 10^{13}$ G in order to explain the observed X-ray luminosity in terms of the release of magnetic energy see \cite{duncan92,thompson95} for details. The inferred upper limit of the surface magnetic field of SGR 0418+5729 $B<7.5\times 10^{12}$ G describing it as a neutron star see Rea et. al.\cite{rea10} for details, is well below the critical field, which has challenged the power mechanism based on magnetic field decay in the magnetar scenario.

Alternatively, it has been recently pointed out how the pioneering works of Morini et. al.\cite{1988ApJ...333..777M} and Paczynski\cite{paczynski90} on the description of 1E 2259+586 as a white dwarf (WD) can be indeed extended to all SGRs and AXPs. These WDs were assumed to have fiducial parameters $M = 1.4 M_\odot$, $R = 10^3$ km, $I = 10^{49}$ g cm$^2$, and magnetic fields $B \gtrsim 10^7$ G see \cite{M2012} for details inferred from the observed rotation periods and spin-down rates. 

The energetics of SGRs and AXPs including their steady emission, glitches, and their subsequent outburst activities have been shown to be powered by the rotational energy of the WD \cite{M2012}. The occurrence of a glitch, the associated sudden shortening of the period, as well as the corresponding gain of rotational energy, can be explained by the release of gravitational energy associated with a sudden contraction and decrease of the moment of inertia of the uniformly rotating WD, consistent with the conservation of their angular momentum. 

By describing SGR 0418+5729 as a WD, we calculated \cite{M2012} an upper limit for the magnetic field $B < 7.5 \times 10^8$ G and show that the X-ray luminosity observed from SGR 0418+5729 can be well explained as originating from the loss of rotational energy of the WD leading to a theoretical prediction for the spin-down rate
\begin{equation}\label{eq:Pdotnew}
\frac{L_X P^3}{4\pi^2 I} = 1.18\times 10^{-16} \leq \dot{P}_{\rm SGR 0418+5729} < 6.0\times 10^{-15} \, ,
\end{equation}
where the lower limit was established by assuming that the observed X-ray luminosity of SGR 0418+5729 coincides with the rotational energy loss of the WD. As we will show below, these predictions can be still improved by considering realistic WD parameters \cite{kuantay2012} instead of fiducial ones.

The situation has become even more striking considering the recent X-ray timing monitoring with Swift, RXTE, Suzaku, and XMM-Newton satellites of the recently discovered SGR Swift J1822.3--1606 \cite{rea12}. The rotation period $P=8.437$ s, and the spin-down rate $\dot{P}=9.1\times 10^{-14}$ have been obtained. Assuming a NS of fiducial parameters, a magnetic field $B=2.8\times 10^{13}$ G is inferred, which is again in contradiction with a magnetar explanation for this source.

We have recently computed in \cite{kuantay2012} general relativistic configurations of uniformly rotating white dwarfs within Hartle's formalism \cite{1967ApJ...150.1005H}. We have used the relativistic Feynman-Metropolis-Teller equation of state \cite{2011PhRvC..83d5805R} for WD matter, which we have shown generalizes the traditionally used equation of state of Salpeter\cite{1961ApJ...134..669S}. It has been there shown that rotating WDs can be stable up to rotation periods of $\sim 0.28$ s (see \cite{kuantay2012} and Sec.~\ref{sec:3} for details). This range of stable rotation periods for WDs amply covers the observed rotation rates of SGRs and AXPs $P\sim (2$--$12)$ s.

The aim of this article is to describe one of the so-called {\it low magnetic field magnetars}, SGR 0418+5729 as a massive fast rotating highly magnetized WD. In doing so we extend the work of Malheiro et. al.\cite{M2012} by using precise WD parameters recently obtained by Boshkayev et. al.\cite{kuantay2012} for general relativistic uniformly rotating WDs.

%%%%%%%%%%%%%%%%%%%%%%%%%%%%%%%%%%%%%%%%%%%%%%%%%%%%%%%%%%%%%%%%%%
%%%%%%%%%%%%%%%%%%%%%%%%%%%%%%%%%%%%%%%%%%%%%%%%%%%%%%%%%%%%%%%%%%
\section{Rotation powered white dwarfs}\label{sec:2}
%%%%%%%%%%%%%%%%%%%%%%%%%%%%%%%%%%%%%%%%%%%%%%%%%%%%%%%%%%%%%%%%%%
%%%%%%%%%%%%%%%%%%%%%%%%%%%%%%%%%%%%%%%%%%%%%%%%%%%%%%%%%%%%%%%%%%

The loss of rotational energy associated with the spin-down of the WD is given by
\begin{equation}\label{eq:Edot}
\dot{E}_{\rm rot} = -4 \pi^2 I \frac{\dot{P}}{P^3} = -3.95\times 10^{50} I_{49} \frac{\dot{P}}{P^3}\quad {\rm erg s}^{-1}\, ,
\end{equation}
where $I_{49}$ is the moment of inertia of the WD in units of $10^{49}$ g cm$^2$. This rotational energy loss amply justifies the steady X-ray emission of all SGRs and AXPs see \cite{M2012} for details.

The upper limit on the magnetic field see e.g. \cite{ferrari69} obtained by requesting that the rotational energy loss due to the dipole field be smaller than the electromagnetic emission of the magnetic dipole, is given by
\begin{equation}\label{eq:Bmax}
B=\left(\frac{3 c^3}{8 \pi^2} \frac{I}{\bar{R}^6} P \dot{P} \right)^{1/2}=3.2\times 10^{15} \left(\frac{I_{49}}{\bar{R}^6_8}P \dot{P} \right)^{1/2} {\rm G}\, ,
\end{equation}
where $\bar{R}_8$ is the mean radius of the WD in units of $10^8$ cm. The mean radius is given by $\bar{R}=(2 R_{eq}+R_p)/3$ see e.g. \cite{1968ApJ...153..807H} with $R_{eq}$ and $R_p$ the equatorial and polar radius of the star.

It is clear that the specific values of the rotational energy loss and the magnetic field depend on observed parameters, such as $P$ and $\dot{P}$, as well as on model parameters, such as the mass, moment of inertia, and mean radius of the rotating WD.

%%%%%%%%%%%%%%%%%%%%%%%%%%%%%%%%%%%%%%%%%%%%%%%%%%%%%%%%%%%%%%%%%%
%%%%%%%%%%%%%%%%%%%%%%%%%%%%%%%%%%%%%%%%%%%%%%%%%%%%%%%%%%%%%%%%%%
\section{Structure and stability of rotating white dwarfs}\label{sec:3}
%%%%%%%%%%%%%%%%%%%%%%%%%%%%%%%%%%%%%%%%%%%%%%%%%%%%%%%%%%%%%%%%%%
%%%%%%%%%%%%%%%%%%%%%%%%%%%%%%%%%%%%%%%%%%%%%%%%%%%%%%%%%%%%%%%%%%

\begin{figure}
\centering
\includegraphics[width=\columnwidth,clip]{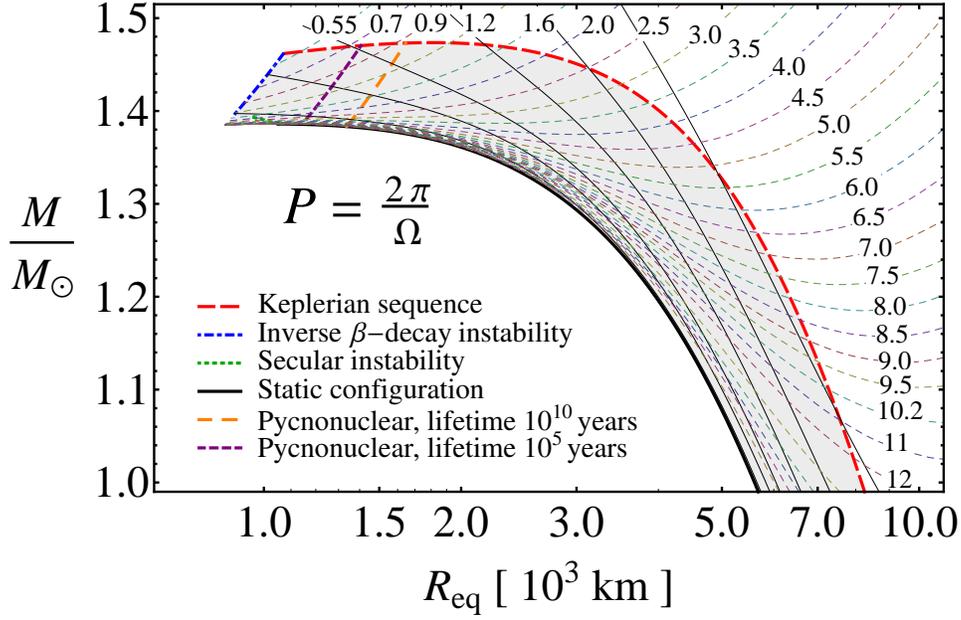}
\caption{Mass versus equatorial radius of rotating Carbon WDs. The solid black curves correspond to $J$=constant sequences, where the static case $J=0$ the thickest one. The color thin-dashed curves correspond to $\Omega$=constant sequences. The Keplerian sequence is the red thick dashed curve, the blue thick dotted-dashed curve is the inverse $\beta$ instability boundary, and the green thick dotted curve is the axisymmetric instability line. The gray-shaded region is the stability region of rotating white dwarfs \cite{kuantay2012}. The numbers show the period $P$ in seconds.}\label{fig:consOmJ}
\end{figure}

The rotational stability of fast rotating WDs was implicitly assumed by Malheiro et. al.\cite{M2012}. The crucial question of whether rotating WDs can or cannot attain rotation periods as short as the ones observed in SGRs and AXPs has been recently addressed by Boshkayev et. al.\cite{kuantay2012}. The properties of uniformly rotating WDs were computed within the framework of general relativity through Hartle's formalism \cite{1967ApJ...150.1005H}. The equation of state for cold WD matter is based on the relativistic Feynman-Metropolis-Teller treatment \cite{2011PhRvC..83d5805R}, which generalizes the equation of state of Salpeter\cite{1961ApJ...134..669S}. The stability of rotating WDs was analyzed taking into account the mass-shedding limit, inverse $\beta$-decay instability, and secular axisymmetric instability, with the latter determined by the turning point method of Friedman et. al.\cite{1988ApJ...325..722F}; see Fig.~\ref{fig:consOmJ} and \cite{kuantay2012}, for details. 

The minimum rotation period $P_{min}$ of WDs is obtained for a configuration rotating along the Keplerian (the mass shedding) sequence, at the critical inverse $\beta$-decay density, namely this is the configuration lying at the crossing point between the mass-shedding and inverse $\beta$-decay boundaries. The numerical values of the minimum rotation period $P_{min}\approx (0.3,0.5,0.7,2.2)$ s and the maximum masses $M_{max}^{J\neq}\approx (1.500, 1.474, 1.467, 1.202)$ were found for Helium (He), Carbon (C), Oxygen (O), and Iron (Fe) WDs, respectively (see Boshkayev et. al.\cite{kuantay2012}, for details). As a byproduct, these values show that indeed all SGRs and AXPs can be described as rotating WDs because their rotation periods are in the range $2 \lesssim P \lesssim 12$ s. 

The relatively long minimum period of rotating Fe WDs, $\sim 2.2$ s, lying just at the lower edge of the observed range of rotation periods of SGRs and AXPs, reveals crucial information on the chemical composition of SGRs and AXPs, namely they are very likely made of elements lighter than Fe, such as C or O.

It can be seen from Fig.~\ref{fig:consOmJ} that every $\Omega$=constant or equivalently $P$=constant sequence intersects the stability region of general relativistic uniformly rotating WDs ($M$-$R_{eq}$ curves inside the shaded region of Fig.~\ref{fig:consOmJ}) in two points. These two points determine the minimum(maximum) and maximum(minimum) $M_{min,max}$($R^{max,min}_{eq}$), respectively, for the stability of a WD with the given rotation angular velocity $\Omega=2 \pi/P$. Associated with the boundary values $M_{min,max}$ and $R^{max,min}_{eq}$, we can obtain the corresponding bounds for the moment of inertia of the WD, $I_{max,min}$, respectively. 

We turn now to a specific analysis of SGR 0418+5729.

%%%%%%%%%%%%%%%%%%%%%%%%%%%%%%%%%%%%%%%%%%%%%%%%%%%%%%%%%%%%%%%%%%
%%%%%%%%%%%%%%%%%%%%%%%%%%%%%%%%%%%%%%%%%%%%%%%%%%%%%%%%%%%%%%%%%%
\section{SGR 0418+5729}\label{sec:4}
%%%%%%%%%%%%%%%%%%%%%%%%%%%%%%%%%%%%%%%%%%%%%%%%%%%%%%%%%%%%%%%%%%
%%%%%%%%%%%%%%%%%%%%%%%%%%%%%%%%%%%%%%%%%%%%%%%%%%%%%%%%%%%%%%%%%%

\subsection{Bounds on the WD parameters}

SGR 0418+5729 has a rotational period of $P=9.08$ s, and the upper limit of the spin-down rate $\dot{P} < 6.0 \times 10^{-15}$ was obtained by Rea et. al.\cite{rea10}. The corresponding rotation angular velocity of the source is $\Omega=2\pi/P=0.69$ rad s$^{-1}$. We show in Table \ref{tab:SGR0418} bounds for the mass, equatorial radius, mean radius, and moment of inertia of SGR 0418+5729 obtained by the request of the rotational stability of the rotating WD, as described in Section \ref{sec:4}, for selected chemical compositions. Hereafter we shall consider general relativistic rotating Carbon WDs.

\begin{table*}
\centering
\tbl{Bounds for the mass $M$ (in units of $M_\odot$), equatorial $R_{eq}$ and mean $\bar{R}$ radius (in units of $10^8$ cm),  moment of inertia $I$, and surface magnetic field $B$ of SGR 0418+5729. $I_{48}$ and $I_{50}$ is the moment of inertia in units of $10^{48}$ and $10^{50}$ g cm$^2$, respectively.}
{\scriptsize
\begin{tabular}{c c c c c c c c c c c}
Comp. & $M_{min}$ & $M_{max}$ & $R^{min}_{eq}$ & $R^{max}_{eq}$ & $\bar{R}_{min}$ & $\bar{R}_{max}$ &
$I^{min}_{48}$ & $I^{max}_{50}$ & $B_{min} (10^7 {\rm G})$ & $B_{max} (10^8 {\rm G})$\\
\hline He & 1.18 & 1.41 & 1.16 & 6.88 & 1.15 & 6.24 & 3.59 & 1.48 & 1.18 & 2.90\\
C & 1.15 & 1.39 & 1.05 & 6.82 & 1.05 & 6.18 & 2.86 & 1.42 & 1.19 & 3.49\\
O & 1.14 & 1.38 & 1.08 & 6.80 & 1.08 & 6.15 & 3.05 & 1.96 & 1.42 & 3.30\\
Fe & 0.92 & 1.11 & 2.21 & 6.36 & 2.21 & 5.75 & 12.9 & 1.01 & 1.25 & 0.80\\
\hline
\end{tabular}
}

\label{tab:SGR0418}
\end{table*}

\subsection{Rotation power and magnetic field}

Introducing the values of $P$ and the upper limit $\dot{P}$ into Eq.~(\ref{eq:Edot}) we obtain an upper limit for the rotational energy loss
\begin{equation}\label{eq:EdotmaxSGR0418}
\dot{E}_{\rm rot} <
\begin{cases}
-9.05\times 10^{32}\quad {\rm erg\,s}^{-1}, & M=M_{max}\\
-4.49\times 10^{34}\quad {\rm erg\,s}^{-1}, & M=M_{min}
\end{cases}\, ,
\end{equation}
which for any possible mass is larger than the observed X-ray luminosity of SGR 0418+5729, $L_X = 6.2\times 10^{31}$ erg s$^{-1}$, assuming a distance of 2 kpc \cite{rea10}.

The corresponding upper limits on the surface magnetic field of SGR 0418+5729, obtained from Eq.~(\ref{eq:Bmax}) are (see also Table \ref{tab:SGR0418})
\begin{equation}\label{eq:BSGR0418}
B <
\begin{cases}
1.19\times 10^{7}\quad {\rm G}, & M=M_{min}\\
3.49\times 10^{8}\quad {\rm G}, & M=M_{max}
\end{cases}\, .
\end{equation}

It is worth noting that the above maximum possible value of the surface magnetic field of SGR 0418+5729 obtained for the maximum possible mass of a WD with rotation period $9.08$ s, $B<3.49\times 10^{8}$ G, is even more stringent and improves the previously value given by \cite{M2012}, $B<7.5\times 10^{8}$ G, based on fiducial WD parameters. 

The electron cyclotron frequency expected from such a magnetic field is 
\begin{equation}\label{eq:fcycSGR0418}
f_{cyc,e}=\frac{e B}{2 \pi m_e c}=
\begin{cases}
3.33\times 10^{13}\quad {\rm Hz}, & M=M_{min}\\
9.76\times 10^{14}\quad {\rm Hz}, & M=M_{max}
\end{cases}\, ,
\end{equation}
corresponding to wavelengths 9.04 and 0.31 $\mu$m, respectively. 

\subsection{Prediction of the spin-down rate}

Assuming that the observed X-ray luminosity of SGR 0418+5729 equals the rotational energy loss $\dot{E}_{\rm rot}$, we obtain the lower limit for the spin-down rate
\begin{equation}\label{eq:PdotminSGR0418}
\dot{P}>\frac{L_X P^3}{4\pi^2 I}=
\begin{cases}
8.28\times 10^{-18}, & M=M_{min}\\
4.11\times 10^{-16}, & M=M_{max}
\end{cases}\, ,
\end{equation}
which in the case of the WD with the maximum possible mass is more stringent than the value reported by Malheiro et. al. \cite{M2012}, $\dot{P}=1.18\times 10^{-16}$, for a massive WD of fiducial parameters.

%%%%%%%%%%%%%%%%%%%%%%%%%%%%%%%%%%%%%%%%%%%%%%%%%%%%%%%%%%%%%%%%%%%
%%%%%%%%%%%%%%%%%%%%%%%%%%%%%%%%%%%%%%%%%%%%%%%%%%%%%%%%%%%%%%%%%%
\section{Concluding Remarks}\label{sec:6}
%%%%%%%%%%%%%%%%%%%%%%%%%%%%%%%%%%%%%%%%%%%%%%%%%%%%%%%%%%%%%%%%%%
%%%%%%%%%%%%%%%%%%%%%%%%%%%%%%%%%%%%%%%%%%%%%%%%%%%%%%%%%%%%%%%%%%

The recent observations of SGR 0418+5729 \cite{rea10}, $P=9.08$ and $\dot{P} < 6.0 \times 10^{-15}$
challenge the description of these sources as ultramagnetized NSs of the magnetar model of SGRs and AXPs. Based on the recent work of Malheiro et. al.\cite{M2012}, we have shown here that, instead, SGR 0418+5729 is in full agreement with a description based on massive fast rotating highly magnetic WDs. 

From the analysis of the rotational stability of the WD using the results of Boshkayev et. al. \cite{kuantay2012}, we have predicted the WD parameters. In particular, bounds for the mass, radius, moment of inertia, and magnetic field of SGR 0418+5729.

We have improved the theoretical prediction of the lower limit for the spin-down rate of SGR 048+5729, for which only the upper limit, $\dot{P} < 6.0 \times 10^{-15}$ Rea et. al. (2010)\cite{rea10}, is currently known. Based on a WD of fiducial parameters, Malheiro et. al.\cite{M2012} predicted for SGR 0418+5729 the lower limit $\dot{P} > 1.18\times 10^{-16}$. Our present analysis based on realistic general relativistic rotating WDs allows us to improve such a prediction, see Eq.~(\ref{eq:PdotminSGR0418}) for the new numerical values.

We have given in Eq.~(\ref{eq:fcycSGR0418}) an additional prediction of the electron cyclotron frequencies of SGR 0418+5729. The range we have obtained for such frequencies fall into the optical and infrared bands. 

We encourage future observational campaigns from space and ground to verify all the predictions presented in this work.

%%%%%%%%%%%%%%%%%%%%%%%%%%%%%%%%%%%%%%%%%%%%%%%%%%%%%%%%%%%%%%%%
%%%%%%%%%%%%%%%%%%%%%%%% References %%%%%%%%%%%%%%%%%%%%%%%%%%%%
%%%%%%%%%%%%%%%%%%%%%%%%%%%%%%%%%%%%%%%%%%%%%%%%%%%%%%%%%%%%%%%%

\end{document}